# Social Internet of Things: Architectural Approaches and Challenges[1]

Juan Ochoa-Zambrano[1] and Juan Garbajosa[2]

[1] Universidad Politécnica de Madrid, CITSEM, Alan Turing S/N, E-28031, Madrid
js.ochoa@upm.es
[2] Universidad Politécnica de Madrid, ETSISI, CITSEM, Alan Turing S/N, E-28031, Madrid
jgs@etsisi.upm.es

**Abstract.** Social Internet of Things (SIoT) takes a step forward over the traditional Internet of Things (IoT), introducing a new paradigm that combines the concepts of social networks with the IoT, to obtain the benefits of both worlds, as in the case of the Social Internet of Vehicles. With the emergence of the Social Internet of Things, new challenges also arise that need to be analyzed in depth. In this article, the key challenges around the software architecture of the various SIoT system described in the literature are analyzed. One of the conclusions is that SIoT is still at an early stage of development, and therefore, SIoT systems architecture will be concerned by this fact. Challenging quality attributes specific for SIoT include scalability, navigability and trust.

## 1 Introduction

Social IoT, in short SIoT, has been introduced as a new paradigm to describe the convergence of social networks and IoT to set up social interaction in an autonomous way according to the rules set by an owner[1]. This paradigm implements an ecosystem facilitating the interaction of people and smart devices. This interaction takes place within a social structure of relationships resembling traditional Social Network Services [2]. The benefit of SIoT over traditional IoT, is that the generated relationships help smart objects learn about other either homogeneous or heterogeneous, objects in a distributed and autonomous fashion [3]. In fact, each device or object can search for help or a specific service by using its relationships, finding friends, and friends of friends by using a query method in a distributed manner [4]. The number of objects connected to the internet has grown considerably during the last few years [5]. According to CISCO [6] around 2.5 quintillion bytes of data are generated every day and, by 2020, around

[1] Work partially funded by the CROWDSAVING project (Ref. TIN2016-79726-C2-1-R).

50 billion objects or devices will be connected to Internet. At this point, SIoT will probably boost the capability of the IoT devices to select, discover, and comprise different services and information gathered by the heterogeneous devices connected to the physical (Internet) world [7].

From the beginnings of SIoT [8], this means that we go back to the year 2011, many architectures have been proposed, implemented and evaluated. Several applications domains such as the Internet of Vehicles in the form of the Social Internet of Vehicles (SIoV) [9][10], the E-health sector [11], recommendation systems[3], for defining human behavior[12], and mobile crowdsensing [13] have exemplified these architectures. However, SIoT has not yet achieved maturity, and many challenges have to be tackled. Some of them are architectural design, services, management, interoperability, implementation, operation and maintenance, scalability, navigability, application development, socio-technical networking, privacy, trustworthiness, and security, fault tolerance interaction and interfaces[14].This paper analyzed the available literature to determine challenges related to the various SIoT systems software architecture available described in the literature.

The remainder of the paper is organized as follows: In Section 2, we delve into the definition of the Social Internet of things, the components and the relations between objects. Some research methodological guidelines followed to identify those challenges related to software architecture are described in Section 3. These challenges are presented in Section 4, through the analysis of the key quality attributes for SIoT, and in Section 5, through the various SIoT architectures and components. In section 6 we discuss the findings from analyzing attributes and architectures. Conclusions and future research guidelines are provided in Section 7.

## 2 Background

In analogy with the traditional Social Networks Services for human beings, the SIoT is a paradigm that introduces the concept of social relationships among objects [8] [15] [16] [17]. Humans usually interact with each other in a wide variety of relationships during everyday life [18]. Likewise, intelligent objects can imitate this behavior establishing new relationships to find and exploit a service as needed. Each object will be the bearer of its specific service to the community; and objects can condition their relationship of *''friendship''* to the achievement of mutual benefits [1]. In that way, a new ecosystem emerges from the clustering of social networks and IoT, enriching both paradigms, since IoT provides the connection to the physical world by sensing and actuating, while social networks contribute with many of the daily aspects of the human world [18].

In SIoT every node is an object capable of establishing social relationships with other objects (i.e. things), in an autonomous way with respect to its owner, capable to solve problems of network navigability and service discovery when the IoT is made of huge numbers of heterogeneous nodes[19]. Therefore, every object can then interact with its friends when needing for some assistance, such as the provisioning of a piece of important information or a key service [5].

SIoT can be a key component to integrate ubiquitous computing in our future daily life. Before this happens, it is necessary to improve the *connectivity* of all relationships between users and things [18]. There is also a clear need to cope with how the concept of interconnecting objects can influence human lives and, therefore, to understand how SIoT can play a role, probably vital, in our *smart* society [12].

The SIoT paradigm presents several good points [1][8]:

- Provides IoT with a social structure to ensure the network navigability and service discovery.
- The degree of interaction among *"friends"* things can be managed through levels of trustworthiness.
- It can be addressed through models traditionally designed to study social networks.

During last years, different types of relationships between objects have been defined to instantiate the essential concept of SIoT, that is, the creation of social relations. Some of these are proposed in [1][8], and are listed below:

- Parental object relationship: established between objects belonging to the same manufacturer.
- Ownership object relationship: established between heterogeneous objects, which belong to the same user.
- Co-location object relationship: established between heterogeneous and homogeneous objects used always stay in the same location.
- Co-work object relationship: established whenever objects collaborate to provide a common IoT application.
- Social object relationship: established when objects come into contact, sporadically or continuously, due to the interactions of their owners.

Other study [20] classifies the previous relationships into two main groups: *profiling relationships* and *dynamic relationships*. The *profiling relationships* are based only on the profile information of objects, and do not depend on the owner behavior. *Ownership object relationship*, *co-location object relationship* and *parental object relationship* belong to this category [20]. Dynamic relationships are, for example, *co-work object relationship* and *social object relationship* [20]; they are created when users, and consequently objects, interact with each other satisfying a number of rules [8]. Study [9] introduces a new interesting relationship, called *guardian object relationship*, part of the Social Internet of Vehicles; in this scenario *vehicles on-board units* become children of *road side units* super nodes [21]. Authors in [22] define three new relationships between devices: *kinship* for the same model of devices from the same manufacturer, *thriendship* for the relationship between devices owned by friends and *shared ownership* for the devices owned by the same user [22].

Most recent works extend the relationships proposed in [8], adding the *sibling object relationship,* created between objects that belong to a group of friends or family members, the *guest object relationship,* which is formed between objects belonging to users acting as guests, the *stranger object relation,* that applies when an object encounter the presence of other object in an anonymous environment such as on the go or in the public environment, and finally the *service object relationship,* that appears when objects form a relationship while coordinating in the same service composition to fulfill a service request [21].

Relationships are the principal component of SIoT. The selection of the right friend among the potential candidates becomes one of the key factors that influences the overall system performance [23]; in fact, the entire system performance, scalability, navigability, safety, and reliability depend on this.

## 3 Methodology

The study started from a research question: Which are the essential elements of the SIoT systems architecture? This question was answered performing a systematic mapping of the available literature. First, Scopus[2] and WebOfSicence[3] libraries were searched through, with the following string: "Social Internet of Things" OR "SIoT". The articles selected for this study include publications from 2011 (the year in which the term SIoT [8] was first published), until the first quarter of 2019. The first filter applied was the title, and then proceeded with the abstract of the selected articles. With a reduced number of selected articles, a full reading of these was made, to then carry out a reverse and forward snowball process. In this way, 52 relevant articles were identified, from which the architectures, characteristics and components for the SIoT were extracted.

## 4 Quality Attributes for SIoT

The key quality attributes for SIot were identified. Table 1 shows those articles that qualifies an attribute as essential or relevant.

**Table 1.** Quality attributes addressed in the literature

| Quality Attribute | Where addressed |
|---|---|
| Interoperability | [3][18][24] |
| Heterogeneity | [18][25] |
| Trust and Trustworthiness | [2][3][4][8][9][10][13][14][15][16][17][26] |
| Security | [2][3][4][6][7][8][9][10][11] |
| Privacy | [3][4][6][7][9][10] |
| Scalability | [4][5][15][20][21][22] |
| Navigability | [3][4][12][19][20][22][23][24] |
| Fault Tolerance | [18] |
| Distribution | [24][25] |

2 Scopus: https://www.scopus.com/
3 Web of Science: https://apps.webofknowledge.com/

Although many attributes were identified, some are inherent in the IoT itself: interoperability, heterogeneity, fault tolerance, privacy, and security [27] [28]. In the current study, we will focus on those attributes that are specific to SIoT, and key for the establishment of social relationships between objects. These attributes are the scalability, navigability and the trust. These attributes were also the most popular, considering the number of studies.

### 4.1 Friendship: Navigability and Scalability

As can it be inferred from the earlier section, the creation of relationships and therefore the selection of friends is the fundamental factor behind the SIoT. In fact, the selection of the right friends can improve the overall network navigability and scalability [3]. Social relationships are closely related to network navigability [1]. SIoT is composed of a large number of objects and each object normally maintains a large number of friends which potentially slow down any kind of search operations [3]. Even the simplest fact of finding the best friend from a set of friends influence the overall system performance and the computational cost [23]. Each object is capable of establishing social relationships with other objects in an autonomous way with respect to its owner; services discovery when establishing social relationships turns out problems when SIoT is made of huge numbers of heterogeneous nodes[19].

The number of relationships that an object establishes [23] is an important variable/indicator. Scalability is important to implement searching and finding the right object that will provide a desired service [29]. To profit from scalability mechanisms, it is crucial to respect the friend selection agreed policy. Friend selection policies, a-priori rules to establish social ties, are a way to avoid the need for central controllers [23]. Thus, the key components that affect service discovery or composition are network navigability and the scalability; and, that also depend on the selection of the right friend.

### 4.2 Trust and Trustworthiness

Trust can be defined as the degree to which a user or other stakeholder has confidence that a product or system will behave as intended [30]. Trustworthy data refers to data and related information that is accurate, complete, relevant, readily understood by and available to those authorized users who need it to complete a task [31]. In service oriented applications, service consumer decisions regarding the service providers depend on the evaluation of the trustworthiness of a service provider along the social trust paths between [32] the consumer and provider. The management of trust between several entities is implemented (1) by providing methods to define trust between the entities and (2) by determining the trustworthiness of the other entity through an automated mechanism [32]. In a certain way, to establish a degree of trustworthiness in SIoT, the degree of interaction or relationships among things [1] [25] can be leveraged.

SIoT depends upon the relationship and trustworthiness between a trustee and a trustor[33]. In that way, trust plays an important role in selecting the right friends, consulting the appropriate service provider, and evaluating the trustworthiness of the peers and community[23]. Then, the main aim of trustworthiness is understanding how the information provided by other members or objects has to be processed[34].

Clearly, the performance of such a kind of process in the SIoT network is strictly subject to the capability of the objects to replicate the human innate behavior in handling social relationships, like reputation to measure the trustworthiness of an object[26]. Trust is not only a property of a trustor or a trustee, but it is also a relationship between the trustor and the trustee that is subjective and asymmetric, derived from the triad of trustee's trustworthiness, trustor's propensity, and environment's characteristics[35]. Trustworthiness of transmission, of sensed data, and of computing results have been identified as challenges that must be addressed [36].

## 5 SIoT Architectures and components

This section will address, first the SIoT architectures that are described in the literature, and, second, will study more in detail some of the architectural components. Using the systematic mapping reported in Section 3, 20 articles were identified as presenting new architectures, applications or implementations for SIoT.

### 5.1 Sample Architectures for SIoT

Aztori et al. [8] presented a three-layered architecture: **(1) Object layer,** that contains the physical objects and their specific communication interfaces. **(2) Component layer** includes components as Identification Management, Object profiling, Owner control, Relationship management to handle all the relationships, Service discovery for finding an object that provides a required service, Service composition to enables the interaction between objects, and Trustworthiness management to understand how the provided information must be processed**. (3) Application layer** includes Human interfaces, Object Interfaces and Service API components.

A variation of the previous architecture is introduced in [1], this architecture is composed of: **(1) SIoT server**, **(2) Gateway layer** and **(3) Object layer**. The tree-layered architectures proposed in [8] are present in this architecture as a single component called SIoT Server, that contains the Application and Network layer [1]. Also, a Gateway and Object architectures were introduced, both of them containing three layers [1]. A point to emphasize is that the basic structure of the Component layer proposed in [8] remains the same in [1].

An interesting application domain is the Internet of Vehicles (IoV). The authors of reference [9] applied SIoT to the Internet of vehicles, and provided an architecture for the so called SIoV. The architecture was composed of six elements: **(1) Home Base Unit:** Identity Manager, Data Manager, Dispatcher, Privacy settings**. (2) On Board**

**Unit**: Identity Manager, Automotive Ontology, Message Builder, Data Manager, Message Builder and Dispatcher. **(3) tNote Message:** automotive ontology [37], Dedicated Short Range Communications and Advanced Traveler Information System. **(4) Road Side Unit:** Identity Manager, Data Manager, Dispatcher, Social Tag Manager. **(5) tNote Cloud:** Topology Optimizer, Query Processor and Data Manager. **(6) User Interface:** Profile, Routes, Friends, Groups, and Social Graph.

Alam et al. [10] applied the architecture proposed in [9] , and reducing it to three main components: **(1) Physical Entities**, **(2) Cyber Entity**, and **(3) Social Graph.** There is no change in the essential components of the architecture.

An alternative and the most recent architecture for SIoV can be found in [24]. This architecture has the following layers**: (1) Physical world layer**: Vehicles, environmental sensor **(2) Gateway layer**: Smart vehicle module, roadside unit **(3) Fog layer**: fog node components **(4) Cloud layer**: resources, analytics, and big dat**a (5) Application layer**: applications, services **(6) Users layer**: drivers, passengers, pedestrians, and ITS agents. All the aforementioned layers are governed by the Trust Manager, Social Relationship Manager and Security and privacy module[24].

Reference [19] introduced a three-layered architecture: (1) **Sensing layer**: information from physical world, (2) **Communication layer**: communication networks, and **(3) Application layer**: services.

The authors of [38] proposed an architecture based on ontologies consisting of three layers: **(1) Communication layer:** Semantic Restful Client and Third-party Restful API**. (2) Control layer:** Context Handler, Command Executor, Ontology and User Rules**. (3) Ontology-based Layer:** Recommendation Reasoner, Profile and Rules handler. In this architecture, the system's ontology database plays a crucial role, since it defines information classes and their relationships [38].

The authors of [36] introduced the SIoT paradigm into Crowdsourcing, using the same architecture proposed in [8]. The sensing entities of the crowdsourcing participants have an **(1) Entity layer**, and **(2) Abstraction layer**, and a **(3) Social proxy** [36].

An interesting application of SIoT is the introduced in [12], that combines the Big Data with SIoT to define human behaviors. The proposed big data analytics ecosystem consists of three domains[12]: **(1) Object domain**: traditional IoT devices and social networks. **(2) SIoT server domain:** load balancer, data storage, ID management, preprocessing and processing modules. **(3) Application domain:** security, cloud server, result storage, device and data server.

In [39] "*Lysis platform*" was introduced; it consists of four levels: **(1) Real word level:** Hardware Abstraction Layer, Data Handler, Device Management, Environment Interface/Protocol Adapter**; (2) Virtualization Lever**: Social Enabler, Virtual Object, Social Virtual Object API, Social Virtual Object Hardware Abstraction Level; **(3) Aggregation Level**: composed by Micro Engines; (**4) Application Level**: in which user-oriented macro services are deployed. The main component of this architecture is the SVO(Social Virtual Objects), which is a Virtual Objects[40] plus a Social Enabler. The Social Enabler is composed of Relationship Management, Owner Control, ID Management, Trust Management, Social Virtual Object Search [39]. Some applications of the Lysis [39] platform are shown in [41] and [13], where Lysis was applied to energy efficiency in smart building and mobile crowdsensing respectively.

In [3] a SIoT architecture for Recommendation services was presented, it consists of the next layers[3]: **(1) SIoT perception Layer**: responsible for sensing and collecting information from IoT devices; **(2) Network layer**: composed of various telecommunication networks; **(3) Interoperability Layer:** required for data sharing among various IoT applications due to the different semantics of each IoT application; **(4) SIoT recommendation System**: uses this SIoT data to build and maintain social relationships and profiles between people-and-things, and between things-and-things; **(5) IoT Applications**

An architecture with three components was proposed in [22] structured as follows: **(1) Socialite Server:** Semantic models to manage location and relationships between users and devices, Service Adapter, Device Adapter and Rules. **(2) Databases:** to store persistent data and semantic models. **(3) Socialite client:** End users programming, Remote access and control devices.

In reference [11], an architecture for E-health systems based on the SIoT was introduced. The layers of this architecture are[11]: **(1) Objects plane:** physical world objects**;** **(2) Social objects plane:** smart objects, advanced application hosting device, social gateway and social IoT interface; **(3) Network plane:** SIoT middleware, social object registration, social relationship management, resource discovery and social object mutual authentication; **(4) Virtual Entities plane:** virtual doctor and E-Butler**; (5) User Plane:** interfaces.

An architecture for large-Scale Systems based on Edge computing and SIoT was introduced in [42]. It uses the iSpiens platform [43] to implements SIoT components as follows: relationship management, service discovery, service composition, trustworthiness management. The architectural components are the following: **(1) Physical devices:** physical objects; **(2) In-network edged computation:** virtual object container, social object container; **(3) Off-network computation:** cloud**,** internet services, SIoT platform.

Finally in [21] an architecture of three levels was introduced with **(1) Object virtualization level ; (2) Aggregated object virtualized,** and **(3) Service level**

### 5.2 Key Components

As it can be seen in the previous Section, a large number of architectures have been introduced and many of them have even been fully implemented and assessed. Many of the architectures are complex and made of multiple levels and components, while others simpler. Despite this, it is possible to identify some key components, often common to all the architectures, and that support the attributes outlined in Section 4. The main attributes outlined were navigability, scalability, trust, and trustworthiness, which are directly related with the SIoT social component.

Therefore, the most significant components for a SIoT platform and that makes it different from the traditional IoT, are Service composition, Social Relationship and Trustworthiness Managers. These elements are described below:

- *Service composition*: his component is responsible for generating interactions between different friends. For different tasks, different friends are chosen based on the level of trust or friendship of these partners. The services that can be composed will depend on the established relationships

- *Social Relationship manager*: It is the essential component of the SIoT, being the one in charge of selecting the relations that shall be established.
- *Trustworthiness manager*: is responsible for providing the necessary information on what actions to take with the information received from the different objects.

Within the different presented architectures, we can also find other common elements such as the physical layer or the real world layer, where we can find the objects, and the communication layer that enables interaction between them. These layers are not of interest for SIoT, since these have been already studied within conventional IoT.

## 6 Analysis and discussion

Since the emergence of the term *SIoT*, many architectures have been introduced, and a number of promising applications have been produced deploying them. Therefore, the potential of the SIoT over the traditional IoT has been showed. Section 5 shows that all these architectures, considering also the various applications domains, are made of components that share, often, similar functions. These components bring about new challenges that should be carefully studied.

The concepts of social networks into IoT to devise SIoT, enables IoT objects to autonomously establish relationships with other objects so that object can find services or can cooperate in a common goal. When devising SIoT navigability and scalability of the network were also considered as goals. However, the complexity of the relationships that an object can create and the selection of the right friends turn out new issues that will directly impact navigability and scalability[18].

The selection of the right friends is a crucial factor that directly impacts the performance of the entire platform[23], and specially the system navigability and scalability. The design of the Relationship Manager may be critical for the platform performance.

Trust is a factor that not only has an impact on the choice of a potential friend, but also directly on the service discovery and composition [44]. In fact, since an object can have a large list of friends, trust will be a factor that will help to decide which the right friend to consume a certain service is. Choosing the right friend has a special impact on the system reliability.

From the reviewed articles, it has been found that Trust has been extensively addressed and also how this factor helps establish the social relations between the objects, however, though more work might be needed to define how Trust may change from the moment before establishing a Friendship to after this happens. For instance, consider the case of establishing a new relationship; although there are certain types of friendship already described in the literature, it is still necessary to define the level of friendship more in depth, just as in real life happens: human beings have school friends, close friends, best friends, etc. That is, for different types of activities, humans choose different friends, similarly, different friends or objects must be selected in SIoT to address different types of tasks.

At this point, it is clear that the SIoT key element is the establishing of social relationships; but first it is necessary to establish better criteria for the categorization, selection and evaluation of friends. Since the correct selection of friends affects the navigability and scalability of the system, it is necessary to answer the following questions first. How many friends are enough? Is it better to have many good friends? Alternatively, is a balanced group of good and bad friends better?

Another important point falls on the Service Composition. Several articles propose applying SIoT to the E-health sector [11], to recommending systems [3], to defining human behavior[12], or to mobile crowdsensing [13]. Ideally, instead of proposing a new component for each new application or service, these tasks must be directly taken on by the Service Composition [1] [8] or the Micro Engines [39]. In this way, a unique architecture for Service Composition could be developed to take care of all kind of intelligence tasks. The Service Composition would be the component in charge of choosing the objects or friends that will be used for a certain task, for example, in the case of Crowdsourcing activities: it could make a selection of heterogeneous groups of good and bad friends, while for Recommendation tasks it could choose a more homogenous group only made of friends with good reputation.

Therefore, more research is needed to study how and which friendship relations are to be established, how to use them, and how Trust will be changing throughout a relationship, as well as determining what are the mechanisms used to establish the amount and the correct type of friends for each task.

As a final remark, it can be said that the essential elements of SIoT need to be further worked and studied because there are still many issues to solve, for future interactive and collaborative platforms that include humans, not only machines.

## 7 Conclusions and future work

This paper has analyzed SIoT, mainly from an architectural perspective, but addressing not only structural issues but also architectural qualities. It is clear that SIoT opens new perspectives, with respect to IoT.

SIoT is still at an early stage of development. In fact, basic issues such as the significance of trustworthiness, and the implications of this, for instance in the case of Internet of Vehicles, makes it necessary to rethink how other qualities are ensured. Also basic and fundamental, it is necessary to improve how the selection and leveraging of friends are performed. For different tasks, different types of friends will be required. And, it is still necessary to determine the right number of friends for the correct operation of the system. Then, it is necessary to evaluate if the diversity of friends can improve or impact the performance of the system for certain of the intelligence tasks.

Architectural components, such as the case of the *Relationship manager* or *Service composition*, will require a careful design and much more work. Service composition must be developed more in depth. As SIoT grows as a more mature discipline, more consolidated software architectures will be available.

Looking at the future, it is necessary to focus attention on two directions. First, the theoretical foundations for trust, scalability, and navigability in the context of SIoT must be improved. Second, as well as better design and implementations, it is necessary to enhance indicators that allow us assess qualities. Simply proposing new techniques to improve trust, security, privacy, navigability, scalability and reliability of SIoT-based platforms will not be enough to overcome existing and future challenges.